\documentclass[10pt,conference]{IEEEtran}
\usepackage{amsmath,amssymb,mathrsfs}
\usepackage{epsfig,epsf,subfigure,graphicx,graphics}
\usepackage{url}
\usepackage{color}

\newtheorem{theorem}{Theorem}[section]

\newtheorem{remark}{Remark}[section]
\newtheorem{claim}{Claim}[section]

\newcommand{\lp}{\left(}
\newcommand{\rp}{\right)}

\newcommand{\lbp}{\left\{}
\newcommand{\rbp}{\right\}}

\newcommand{\msf}{\mathsf}

\graphicspath{{../figs/}}

\IEEEoverridecommandlockouts

\allowdisplaybreaks

\newcommand{\dl}{\delta}

\begin{document}

\title{Capacity Region of Erasure Broadcast Channels with Common Message and Feedback}

\author{
\authorblockN{Alireza Vahid}
\authorblockA{
University of Colorado, Denver\\
Department of EE\\
Denver, USA\\
\url{alireza.vahid@ucdenver.edu}} \and
\authorblockN{Shih-Chun Lin}
\authorblockA{
NTUST\\
Department of ECE\\
Taipei, Taiwan\\
\url{sclin@ntust.edu.tw}} \and
\authorblockN{I-Hsiang Wang}
\authorblockA{
National Taiwan University\\
Department of EE\\
Taipei, Taiwan\\
\url{ihwang@ntu.edu.tw}}
}

\maketitle

\begin{abstract}
Jolfaei et al. used feedback to create transmit signals that are simultaneously useful for multiple users in a broadcast channel. Later, Georgiadis and Tassiulas studied erasure broadcast channels with feedback, and  presented the capacity region under certain assumptions. These results provided the fundamental ideas used in communication protocols for networks with delayed channel state information. However, to the best of our knowledge, the capacity region of erasure broadcast channels with feedback and with a common message for both receivers has never been presented. This latter problem shows up as a sub-problem in many multi-terminal communication networks such as the X-Channel, and the two-unicast problem. In this work, we present the capacity region of the two-user erasure broadcast channels with delayed feedback, private messages, and a common message. We consider arbitrary and possibly correlated erasure distributions. We develop new outer-bounds that capture feedback and quantify the impact of delivering a common message on the capacity region. We also propose a transmission strategy that achieves the outer-bounds. Our transmission strategy differs from prior results in that to achieve the capacity, it creates side-information at the weaker user such that the decodability is ensured even if we multicast the common message with a rate higher than its link capacity.
\end{abstract}


\section{Introduction}

The broadcast channel (BC) is one of the first multi-terminal networks studied in Information Theory~\cite{cover1972broadcast}. Recently, this problem with available channel state feedback has attracted more interest~\cite{GeorgiadisDetDelayedBC,Wang_12,GatzianasGeorgiadis_13,maddah2012completely,Jafar14PN} as it plays an essential role in understanding the feedback capacity region of several fundamental problems in network information theory, such as the Interference Channel, the X-Channel, and the two-unicast problem. Of particular interest, is the erasure BC in which each wireless link is on (active) or off (dropped) according to some probability distribution. In a packet-based communication network, each hop can be modeled as a packet erasure channel~\cite{DanaGowaikar_06}, and thus, studying the erasure BCs provides a good understanding of multi-session uni-casting in small wireless networks~\cite{GeorgiadisDetDelayedBC,Wang_12,GatzianasGeorgiadis_13}.

Jolfaei et al.~\cite{jolfaei1993new} leveraged delayed channel state information (CSI) to create transmit signals that are simultaneously useful for multiple users in a broadcast channel. Later, Georgiadis and Tassiulas~\cite{GeorgiadisDetDelayedBC} studied erasure broadcast channels with feedback, and  presented the capacity region under certain assumptions. These results presented the key ideas used in communication protocols for networks with delayed CSI. In~\cite{maddah2012completely}, the auhors showed that the delayed CSI can still be very useful and can change the achievable degrees-of-freedom (DoF). This discovery generated a momentum in studying the DoF region and the approximate capacity region of of multi-terminal wireless networks with delayed CSI. 

Despite all the attention the (erasure) BC has attracted over the decades, we were unable to find the characterization of the capacity region of this problem with feedback when the transmitter has a common message for both receivers as well as a private message for each. This problem naturally arises in more complicated problems such as the X-Channel and the two-unicast problem, highlighting its importance. In this work, we present the capacity region of the two-user erasure BC with delayed feedback, private and common messages. We compare our results to the findings of~\cite{GeorgiadisDetDelayedBC} in Section~\ref{Section:Prior}.  

In particular, we provide a new set of outer-bounds to capture and to quantify the ability of the transmitter in performing interference alignment when channel state is known with delay and when each receiver must be able to decode the common message. These outer-bounds illustrate the impact of delivering a common message on the overall maximum achievable rates. Intuitively, delivering a common message to both receivers reduces the transmitter's ability to perform interference alignment and the achievable region shrinks. In fact, we show that the maximum sum-capacity (including the common rate) is attained when there is no common message, and decreases as the common message rate increases.

The capacity-achieving transmission strategy differs from prior results. One idea is to modify prior results by adding a segment to send the common message after the capacity-achieving scheme for private messages, using an erasure code with the rate corresponding to the weaker receiver. This idea achieves the capacity when the erasure probabilities are equal. However, when erasure probabilities are different, this scheme is no longer optimal. In Section~\ref{Sec_achi}, we will show how to transmit the common message at a higher rate and yet ensure decodability at the weaker receiver. The key is to properly produce side-information of the common message at the weaker receiver during the re-transmission of private bits.


\section{Problem Formulation}
\begin{figure}[!ht]
\centering
\includegraphics[height = 2.5in]{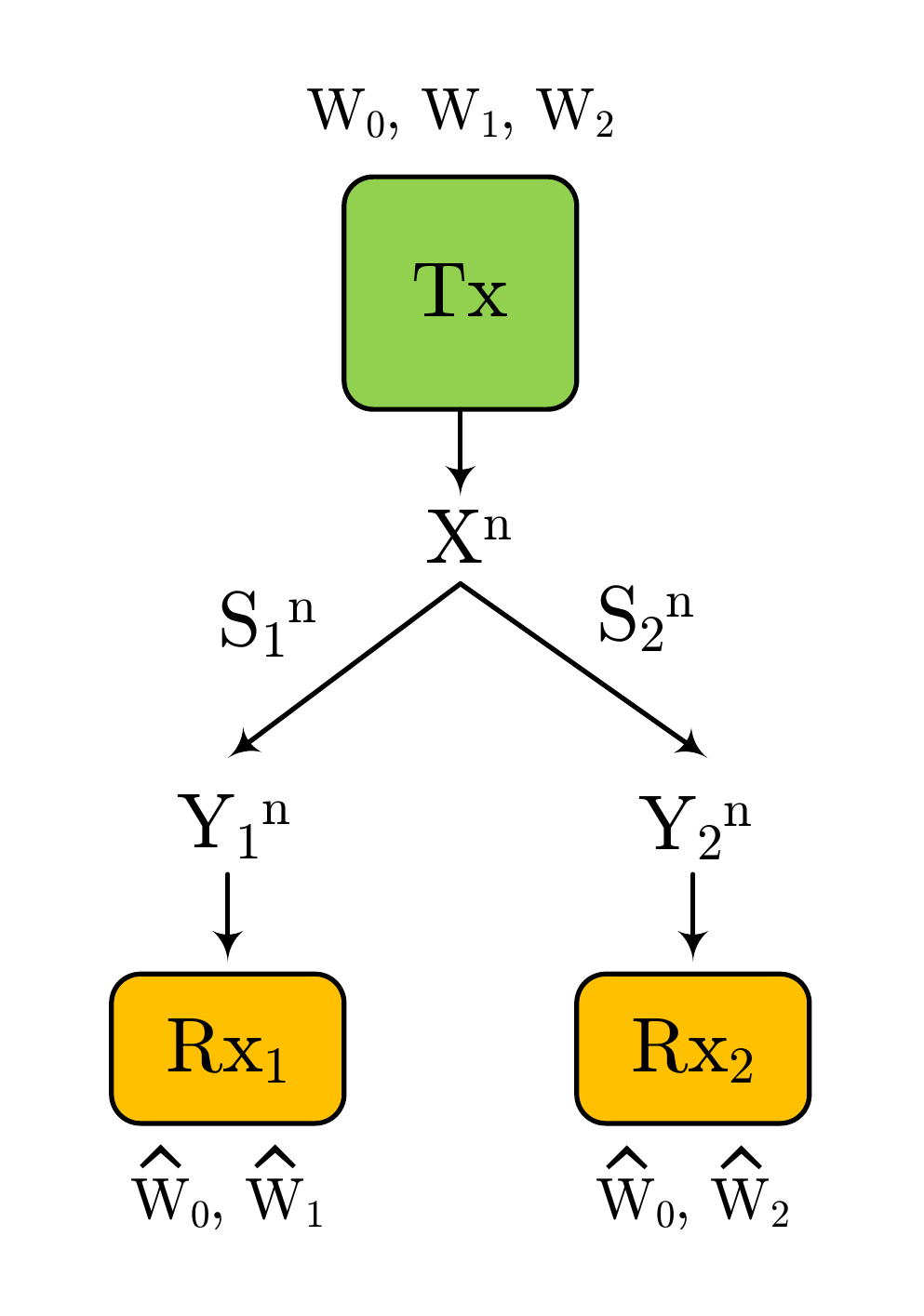}
\caption{Two-user erasure BC with delayed CSI and a common message.}
\label{Fig:BC-Common}
\end{figure}

As in Fig.\ref{Fig:BC-Common}, we consider the two-user binary erasure BC in which one transmitter wishes to transmit messages $W_1$ and $W_2$ to two receiving terminals $\msf{Rx}_1$ and $\msf{Rx}_2$, respectively, as well as one common message $W_0$ to both receivers, over $n$ channel uses. 
Here, $W_i$ is uniformly distributed over $\lbp1,2,\ldots,2^{nR_i}\rbp$, for $i=0,1,2$, and messages are distributed independently. Three messages are mapped to the channel input $X[t] \in \mathbb{F}_2$, the binary field, and the corresponding received signals at $\msf{Rx}_1$ and $\msf{Rx}_2$ respectively are
\begin{align}
Y_1[t] = S_1[t] X[t] \; \mbox{and} \; Y_2[t] = S_2[t] X[t], \label{eq_DL_channel}
\end{align}
where $\lbp S_i[t]\rbp$ denotes the Bernoulli $(1-\delta_i)$ process that governs
the erasure at $\mathsf{Rx}i$, and it is distributed i.i.d. over time. When $S_i[t]=1$, $\mathsf{Rx}i$ receives $X[t]$ noiselessly, and when $S_i[t]=0$, the received signal is mapped to an erasure. We also assume $\delta_{12}=P\{S_1[t]=0,S_2[t]=0\}$.  The messages are distributed independently from the channel realizations.


We assume both receivers feed back their states and thus, the transmitter knows the channel state information (CSI) in scenario ``$\mathrm{DD}$,'' where both $S_1[t]$ and $S_2[t]$ are known with unit delays. The constraint imposed on the encoding function $f_t(.) $ at time index $t$ is
\begin{equation} \label{eq_enc_PID}
X[t] = f_t\lp W_1,W_2, W_0, S_1^{t-1},S_2^{t-1}\rp,
\end{equation}
where $S_i^{t-1}=(S_i[1],\ldots,S_i[t-1]), i=1,2$. We assume that the full CSI, $S^n=(S_1^n,S_2^n)$, is known at each receiver, and the corresponding error probability constraints at $\msf{Rx}i$ is
\begin{align}
  &\Pr\big\{ (W_i,W_0)  \neq g_i ( Y^{[1:n]}_i,S^n)\big\} \rightarrow 0, \notag
\end{align}
as $n \rightarrow \infty$, where $g_i(.)$ is the decoding function
at receiver $i, i=1,2$. The capacity region is the closure of the collection
of all rate triples $(R_1,R_2,R_0)$ satisfying the error probability
constraints.

\section{Main Results}
\label{Section:Main}

Our main result is identifying the capacity region for Fig \ref{Fig:BC-Common}.

\begin{theorem} \label{ThmDDCommon}
For the binary erasure broadcast channel under scenario $\mathrm{DD}$, the capacity region with a common message $W_0$ and private messages $W_1,W_2$, is the collection of all non-negative $(R_1,R_2,R_0)$ satisfying
 \begin{align}
 \frac{R_1}{1-\delta_{12}} + \frac{R_2+R_0}{1-\delta_2} \leq 1, \label{eq_DD_B.1}\\
 \frac{R_1+R_0}{1-\delta_1} + \frac{R_2}{1-\delta_{12}} \leq 1. \label{eq_DD_B.2}
 \end{align}
\end{theorem}
\noindent Without loss of generality, we assume $\delta_1 \leq \delta_2$, and define
\begin{align}
\label{Eq:RBar}
\bar{R} \overset{\triangle}= \frac{(\dl_1-\dl_{12})}{(\dl_2-\dl_{12})}(1-\dl_2).
\end{align}

\begin{figure}[!ht]
\centering
\includegraphics[height = 2in]{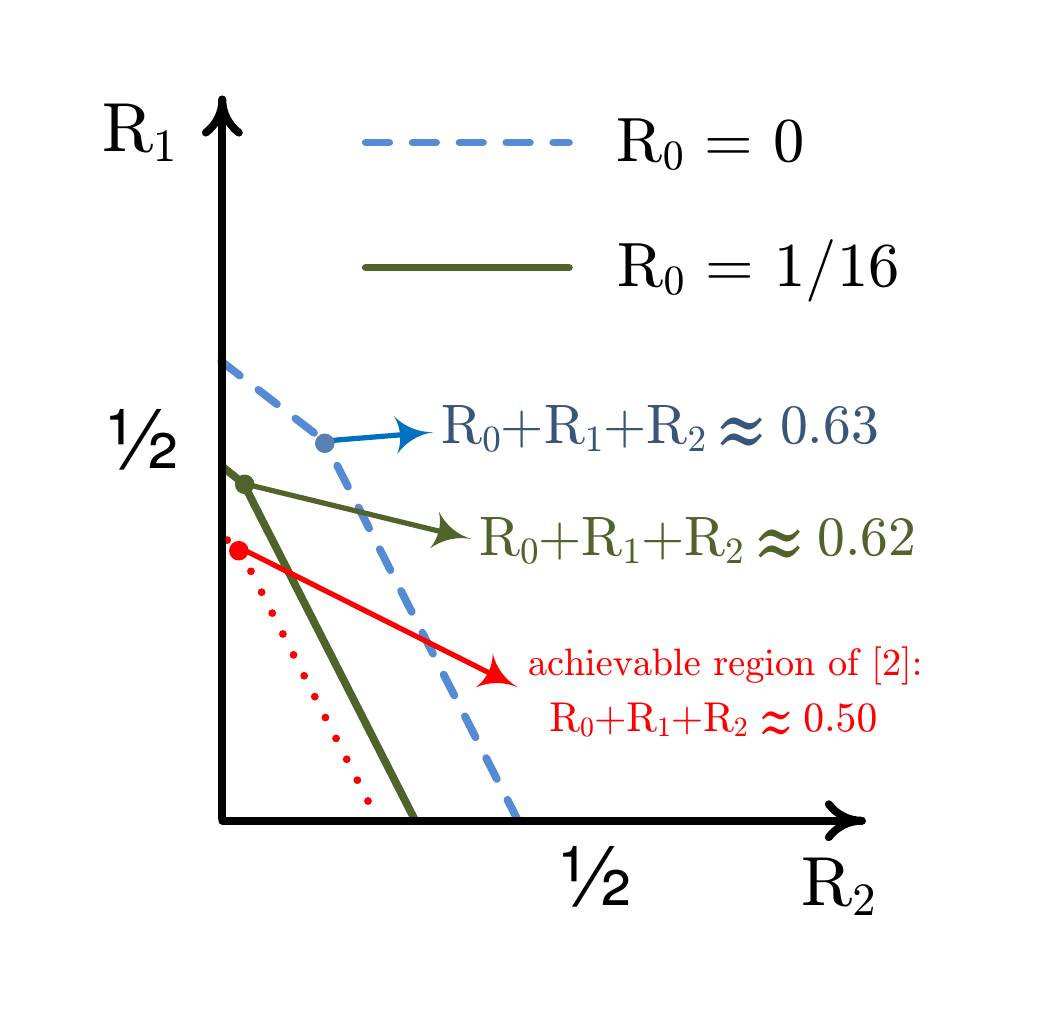}
\caption{Increasing the common rate, $R_0$, decreases the maximum sum rates. The capacity regions, with $R_0<\bar{R}$, are drawn for $\dl_1 = 0.4, \dl_2 = 0.6,$ and $\dl_{12} = \dl_1\dl_2$. This figure also depicts the achievable region of~\cite{GeorgiadisDetDelayedBC} for the given parameters and will be discussed later.} \label{Fig:Region}
\end{figure}

\begin{figure}[!ht]
\centering
\includegraphics[height = 2in]{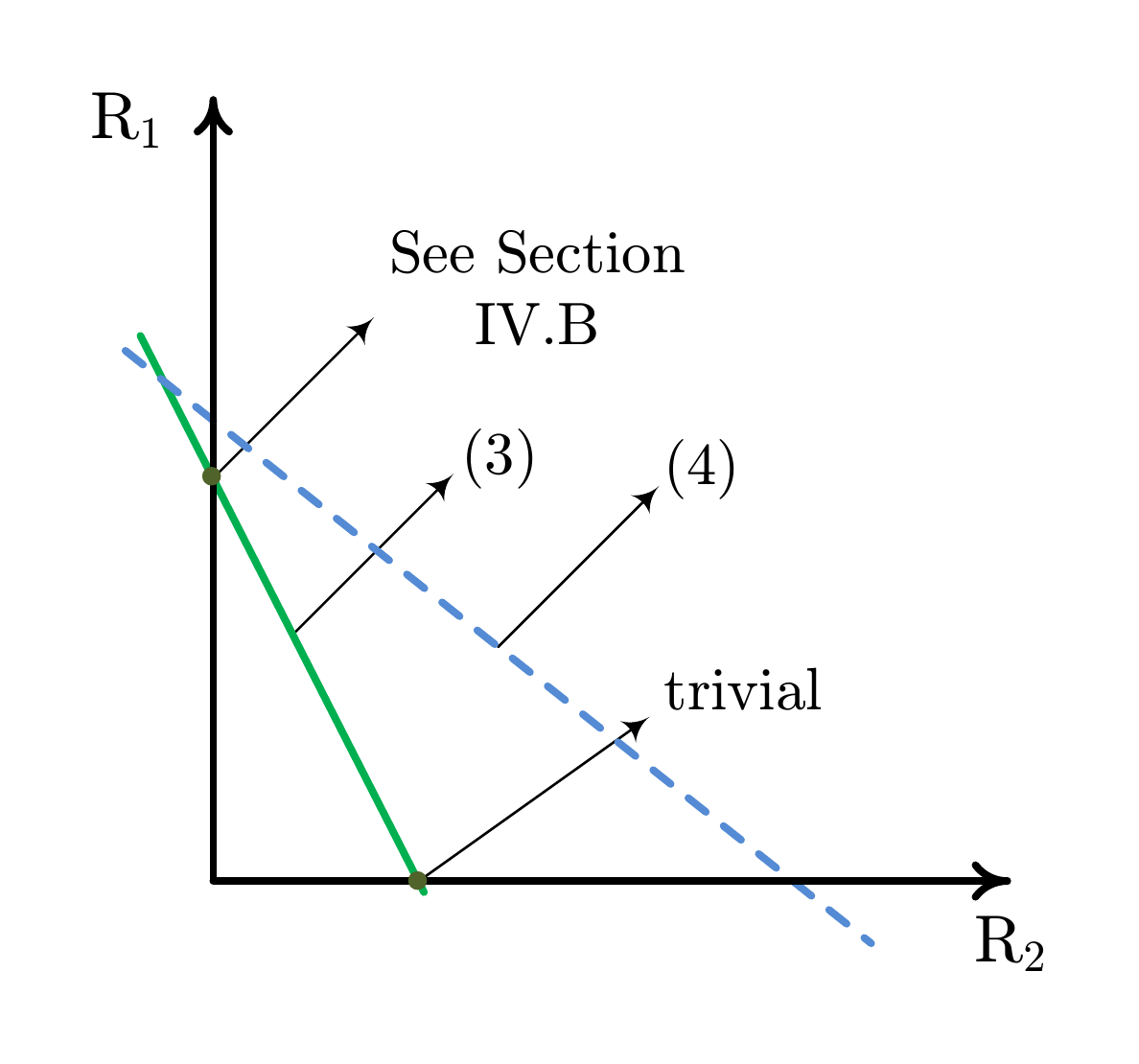}
\caption{Capacity region for $R_0 \geq \bar{R}$. In this case, for $\delta_1 < \delta_2$, only \eqref{eq_DD_B.1} is active. The trivial corner point is $(R_1,R_2) = (0,1-\delta_2-R_0)$.}
\label{Fig:largeR0}
\end{figure}

The capacity region's shape differs whether the common message rate meets $R_0 < \bar{R}$ or not. Fig.~\ref{Fig:Region} illustrates the capacity region when $R_0<\bar{R} \approx 0.178$ with $\delta_1= 0.4,$ $\delta_2=0.6,$ and $\delta_{12}=\delta_{1}\delta_{2}$. Intuitively, delivering more common rate should decrease the overall maximum achievable rates. In Fig.~\ref{Fig:Region} as $R_0$ increases, the region shrinks, and the maximum achievable sum-rate (including the common rate) for $R_0 = 0$ and $R_0 = 1/16$ are approximately $0.63$ and $0.62$, respectively, which represents a decrease. If we further increase $R_0$ such that $R_0 \geq \bar{R}$, the capacity region will have a triangular shape as shown in Fig.~\ref{Fig:largeR0} for $\delta_1 < \delta_2$ where only \eqref{eq_DD_B.1} is active. Appendix~\ref{ImpactofR0} provides more details.

Our converse is the extension of the proof of the outer bound for X-channel in \cite{Vahid18X} to different erasure probabilities $\delta_1 \neq \delta_2$ and correlated links $\delta_{12} \neq \delta_1\delta_2$. The details are given in Section \ref{Sec_converse}. Note that when $R_0 \neq 0$, the converse is not provided in \cite{GeorgiadisDetDelayedBC} or any other prior work to the best of our knowledge. As for the achievability, one can add a segment which uses an erasure code with rate $\min\{1-\delta_1, 1-\delta_2\}$ for sending the common message after the  capacity-achieving re-transmission scheme for only private messages. When $\delta_1=\delta_2$, this simple scheme achieves the capacity region in Theorem \ref{ThmDDCommon}. However, when $\delta_1 \neq \delta_2$, transmitting the common message with rate $\min\{1-\delta_1, 1-\delta_2\}=1-\delta_2$ is no longer optimal. In \ref{Sec_achi}, we will show how to transmit the common message at the higher rate of $(1-\delta_1)$ and yet ensure decodability at the weaker receiver, \emph{i.e.} ${\sf Rx}_2$. The key is to properly produce side-information of the common message at receiver 2 during the re-transmission of private bits.

\section{Comparison to Prior Results of~\cite{GeorgiadisDetDelayedBC}}
\label{Section:Prior}

Authors in~\cite{GeorgiadisDetDelayedBC} present a similar region, see (7) in~\cite{GeorgiadisDetDelayedBC}, to that of Theorem~\ref{ThmDDCommon} of this work. Thus, we ought to compare the two results. There are two issues regarding (7) in~\cite{GeorgiadisDetDelayedBC}. First, this is claimed as an achievable region and no outer-bound is provided. Second, and more importantly, no details of the achievability proof is presented and we believe the claim is flawed. The authors of~\cite{GeorgiadisDetDelayedBC}, only mention: ``In Phase 3 employ linear random coding of packets $K_1^r$, $K_2^r$, $k_{12}$, i.e., include the multicast session packets in the process,'' to support their claim. In the notation of~\cite{GeorgiadisDetDelayedBC},  $K_i^r$ denotes the recycled bits and $k_{12}$ is the common message. As we will show in our achievability, this idea achieves the capacity when the erasure probabilities are equal. However, when erasure probabilities are different, this scheme is no longer optimal. The achievable region based on the scheme of~\cite{GeorgiadisDetDelayedBC} is illustrated in Fig.~\ref{Fig:Region} and is strictly smaller than the capacity region derived in this work. In Section~\ref{Sec_achi}, we will show how to transmit the common message at a higher rate and yet ensure decodability at the weaker receiver. The key is to properly produce side-information of the common message at the weaker receiver during the re-transmission of private bits. See Remark~\ref{Remark:Comparison} in Section~\ref{Sec_achi} for more details.

\section{Proof of Theorem \ref{ThmDDCommon}}
\subsection{Converse} \label{Sec_converse}

To derive~\eqref{eq_DD_B.1}, let $\beta = \left(1-\dl_{12}\right)/ \left(1-\dl_2\right)$.
We have
\begin{align}
n & \left( R_1 + \beta \left\{ R_2 + R_0 \right\} \right) = H\left( W_1 \right) + \beta \left\{ H\left( W_2 \right) + H\left( W_0 \right) \right\} \nonumber \\
& \overset{(a)}= H\left( W_1 | W_0, W_2 \right) + \beta \left\{ H\left( W_2 \right) + H\left( W_0 | W_2 \right) \right\} \nonumber \\
& \overset{(b)}\leq I\left( W_1; Y_1^n | W_0, W_2, S^n \right) + \notag \\ & ~~~~ \beta \left\{ I\left( W_2; Y_2^n | S^n \right) + I\left( W_0; Y_2^n | W_2, S^n \right\} \right) + n \epsilon_n \nonumber \\
& = H\left( Y_1^n | W_0, W_2, S^n \right) + \beta H\left( Y_2^n | S^n \right) \nonumber \\
& ~~ - \beta \left\{ H\left( Y_2^n | W_2, S^n \right) -  I\left( W_0; Y_2^n | W_2, S^n \right) \right\} + n \epsilon_n \nonumber \\
& \overset{(d)}\leq \beta H\left( Y_2^n | S^n \right) + n \epsilon_n \nonumber \\ & \overset{(e)}\leq n \beta (1-\dl_2) + n \epsilon_n = n (1-\dl_{12}) + n \epsilon_n \label{eq_DD_B.11}
\end{align}
where $\epsilon_n \rightarrow 0$ as $n \rightarrow \infty$; $(a)$ follows from the independence of messages; $(b)$ follows from Fano's inequality and messages are independent of channel realizations; $(d)$ follows from Claim~\ref{Claim:LeakageAppendix} below;  $(e)$ holds since
\begin{align}
H\left( Y_2^n | S^n \right) \leq \sum_{t=1}^{n}{H\left( Y_2[t] | S^n \right)} \leq n (1-\dl_2).
\end{align}
Dividing both sides of \eqref{eq_DD_B.11} by $n(1-\dl_{12})$ and letting $n \rightarrow \infty$, we get
\eqref{eq_DD_B.1}. Similarly, we can obtain \eqref{eq_DD_B.2}.

Now, we prove step $(d)$ of \eqref{eq_DD_B.11} in the following claim.
\begin{claim}
\label{Claim:LeakageAppendix}
For the binary erasure broadcast channel under scenario $\mathrm{DD}$ and with a common message $W_0$ and  two private messages $W_1$ and $W_2$,
\begin{align}
\label{Eq:LeakageAppendix}
&H\left( Y_1^n | W_0, W_2, S^n \right) - \notag \\ & \beta \left\{ H\left( Y_2^n | W_2, S^n \right)  -  I\left( W_0; Y_2^n | W_2, S^n \right) \right\} \leq 0.
\end{align}
\end{claim}

\begin{IEEEproof}
Note that proving (\ref{Eq:LeakageAppendix}) is equivalent to proving
\begin{align}
\label{Eq:Leakage2Appendix}
H\left( Y_1^n | W_0, W_2, S^n \right) - \beta H\left( Y_2^n | W_0, W_2, S^n \right) \leq 0
\end{align}
since
\[
H\left( Y_2^n | W_0, W_2, S^n \right) = H\left( Y_2^n | W_2, S^n \right) - I\left( W_0; Y_2^n | W_2, S^n \right).
\]
Then, we have
\begin{align}
& ~~~~H\left( Y_2^n | W_0, W_2, S^n \right) \nonumber \\
& =\sum_{t=1}^{n}{H\left( Y_2[t] | Y_2^{t-1}, W_0, W_2, S^n \right)} \nonumber \\
& \overset{(a)}= \sum_{t=1}^{n}{H\left( Y_2[t] | Y_2^{t-1}, W_0, W_2, S^t \right)} \nonumber \\
& \overset{(b)}= \sum_{t=1}^{n}{(1-\dl_2)H\left( X[t] | Y_2^{t-1}, W_0, W_2, S_{2}[t] = 1, S_{1}[t], S^{t-1} \right)} \nonumber \\
& \overset{(c)}= \sum_{t=1}^{n}{(1-\dl_2)H\left( X[t] | Y_2^{t-1}, W_0, W_2, S^{t} \right)} 
\end{align}
where $(a)$ follows from the fact that signal $Y_2[t]$ at current time $t$ is independent of future channel states; $(b)$ holds since $\Pr\left( S_{2}[t] = 1 \right) = (1 - \dl_2)$;  $(c)$ is true since transmit signal $X[t]$ is independent of current the channel state at time $t$.

Now, we can have \eqref{Eq:Leakage2Appendix} as
\begin{align}
&~~~~\sum_{t=1}^{n}{(1-\dl_2)H\left( X[t] | Y_2^{t-1}, W_0, W_2, S^{t} \right)} \nonumber \\
& \overset{(d)}\geq \sum_{t=1}^{n}{(1-\dl_2)H\left( X[t] | Y_1^{t-1}, Y_2^{t-1}, W_0, W_2, S^{t} \right)} \nonumber \\
& \overset{(e)}= \sum_{t=1}^{n}{\frac{(1-\dl_2)}{(1-\dl_{12})}H\left( Y_1[t], Y_2[t] | Y_1^{t-1}, Y_2^{t-1}, W_0, W_2, S^{t} \right)} \nonumber \\
& \overset{(f)}= \sum_{t=1}^{n}{\frac{(1-\dl_2)}{(1-\dl_{12})}H\left( Y_1[t], Y_2[t] | Y_1^{t-1}, Y_2^{t-1}, W_0, W_2, S^{n} \right)} \nonumber \\
& \overset{(g)}= \frac{(1-\dl_2)}{(1-\dl_{12})}H\left( Y_1^n, Y_2^n | W_0, W_2, S^{n} \right) \nonumber \\
& \geq \frac{1}{\beta} H\left( Y_1^n | W_0, W_2, S^{n} \right),  
\end{align}
 where $(d)$ holds since conditioning reduces entropy; $(e)$ holds since $\Pr\left( S_{1}[t] = S_{2}[t] = 0 \right) = \dl_{12}$; $(f)$ is true since all signals at current time $t$ are independent of future channel states; $(g)$ follows from the chain rule and the final lower-bound comes from the non-negativity of the entropy function for discrete random variables.
\end{IEEEproof}

\subsection{Achievability} \label{Sec_achi}
We divide the achievability into two cases according to whether or not the common message rate $R_0$ is larger than $\bar{R}$ defined in (\ref{Eq:RBar}). We will illustrate the simpler case $R_0 > \bar{R}$ first, and then the other case, \emph{i.e.} $R_0 \leq \bar{R}$.

\noindent {\bf Case I $R_0 > \bar{R}$}: In this case, as shown in Fig.~\ref{Fig:largeR0}, one can easily check that only outer-bound \eqref{eq_DD_B.1} is active. From \eqref{eq_DD_B.1}, the corner point $R_1=0, R_2=(1-\delta_2)(1-R_0)$ can be trivially achieved by time sharing between the codewords for private message $W_2$ and common message $W_0$. Thus, we focus on the non-trivial corner point given by
\begin{align}
R^*_1 = (1-\dl_{12})\left( 1 - \frac{R_0}{(1-\dl_2)} \right), \qquad R^*_2 = 0. \label{eq_DD_corner_lareR0}
\end{align}
To achieve this point, we allocate $k_1$ private bits for user 1 and $k_0 = (R_0/R^*_1) k_1$ common bits for both users. Then, we adopt the following transmission scheme.

\noindent \textbf{Phase 1:} Send the $k_1$ private bits for user 1 in
\begin{align} \label{eq_DD_T1}
n_1=\frac{k_1}{1-\delta_{12}}
\end{align}
time slots, and we have
\begin{align} \label{eq_DDk12}
k_{1|2} = \frac{(\delta_1-\delta_{12})}{1-\delta_{12}}k_1
\end{align}
bits that are mis-sent to receiver 2 and needed at receiver 1.

\begin{remark}
To keep the description of the protocol simple, we use the expected value of the number of bits in different states, \emph{e.g.}, (\ref{eq_DDk12}). A more precise statement would use a concentration theorem result such as the Bernstein inequality to show the omitted terms do not affect the overall result and the achievable rates~\cite{MaddahAliTIT15}. If at any point the number of bits is not an integer number, we can use $\lceil \cdot \rceil$, the ceiling function, and the results remain unaffected in the limit. 
\end{remark}

\noindent \textbf{Phase 2:} In this phase, we have two segments, namely Segment a and b. \\
\noindent \underline{Phase 2, Segment a}: Encode $k_{1|2}$ mis-sent private bits to receiver 1 using erasure code with rate $1 - \delta_1$ (rate $(1-\dl_1)$ linear code with each entry of its generator matrix randomly generated from an i.i.d. Bernoulli random variable with parameter $1/2$), which takes 
\begin{align} \label{eq_T2a_largeR0}
n_{2a} = \frac{(\delta_1-\delta_{12})}{(1 - \delta_1)(1-\delta_{12})}k_1
\end{align}
time slots. At the same time encode
\begin{align} \label{eq_k02a_largeR0}
k_{0,2a} = n_{2a}(1-\delta_2) = \frac{(1-\dl_2)(\delta_1-\delta_{12})}{(1 - \delta_1)(1-\delta_{12})}k_1
\end{align}
common bits using erasure code with rate $(1-\dl_2)$. Send the XOR of above two encoded sequences.

  \noindent \underline{Phase 2, Segment b}: In this segment, we encode all $k_0$ common bits using random linear codes with length
  \begin{equation} \label{eq_T2b_largeR0}
  n_{2b} = \frac{k_0}{(1-\dl_2)} - n_{2a},
  \end{equation}
and send the encoded bits. Note that $n_{2b}$ is shorter than $k_0/(1-\dl_2)$,  and compared with the simple scheme mentioned at the end of Section~\ref{Section:Main}, the transmission rate at this segment is higher than $\min \{1-\dl_1, 1-\dl_2\} = 1-\dl_2$.

  \noindent \textbf{Achievable rate calculation}:
  First note that the mis-sent $k_{1|2}$ private bits are already known at receiver 2 in Phase 1, and thus, receiver 2 gets $k_{0,2a}=n_{2a}(1-\delta_2)$ common bits at the end of Phase 2, Segment a. Also with the help of these $k_{0,2a}$ bits, receiver 2 can successfully decode all $k_0$ common bits from $n_{2b}(1-\delta_2)$ received bits in Phase 2, Segment b.

Receiver 1 first decodes the common message. To ensure correct decoding of all $k_0$ common bits at receiver 1, we need
  	\begin{equation} \label{eq_hatR1_largeR0}
	\frac{k_0}{n_{2b}} = (1-\dl_2) +  \frac{k_{0,2a}}{n_{2b}} < (1-\dl_1),
	  \end{equation}
where the equality comes from \eqref{eq_k02a_largeR0} and \eqref{eq_T2b_largeR0}, and the inequality is verified in Appendix \ref{proof_hatR1_largeR0}. After removing the interference resulting from the common message at receiver 1 during Segment a, all mis-sent $k_{1|2}$ private bits can be decoded. Then, user 1 will be able decode its private message.

Finally, we calculate the achievable rate of the aforementioned scheme. The total time slots needed are
 \begin{align}
 n &= n_1 + n_{2a} + \frac{k_0}{(1-\dl_2)} - n_{2a} \nonumber \\
 &= \frac{k_1}{1-\delta_{12}} + \frac{k_0}{(1-\dl_2)} \nonumber \\
 &= \frac{k_1}{1-\delta_{12}} + \frac{k_1}{(1-\dl_2)} \frac{R_0}{(1-\dl_{12})(1-\frac{R_0}{(1-\dl_2)})} \nonumber \\
 &= \frac{(1-\dl_2)k_1}{(1-\delta_{12})(1-\dl_2-R_0)}, 
 \end{align}
 where the third equality comes from $k_0=(R_0/R^*_1)k_1$ and \eqref{eq_DD_corner_lareR0}. It can be easily checked that the achievable private rate $k_1/n$ equals the target $R^*_1$, and thus, $k_0/n=R_0$.


 \noindent {\bf Case II $R_0 \leq \bar{R}$}: In this case, both outer-bounds \eqref{eq_DD_B.1} and \eqref{eq_DD_B.2} are active, and the maximum sum-rate corner point is
 \begin{align}
 R^\ast_1 &= \frac{(1-\dl_1)(\dl_2-\dl_{12})-(\dl_1-\dl_{12})R_0}{(1-\dl_{12})-\frac{(1-\dl_1)(1-\dl_2)}{(1-\dl_{12})}}, \nonumber \\
 R^\ast_2 &= \frac{(\dl_1-\dl_{12})(1-\dl_2)-(\dl_2-\dl_{12})R_0}{(1-\dl_{12})-\frac{(1-\dl_1)(1-\dl_2)}{(1-\dl_{12})}}. \label{eq_DD_corner_smallR0}
 \end{align}
 Note that since $\dl_2 \geq \dl_1$, we have $R^\ast_1 \geq R^\ast_2$ from $(1-\dl_1)(\dl_2 - \dl_{12}) \geq (1-\dl_2)(\dl_1-\dl_{12})$ and $-R_0(\dl_1-\dl_{12}) \geq -R_0(\dl_2-\dl_{12})$.

  To achieve the corner point in~\eqref{eq_DD_corner_smallR0}, we fix $k_1$ private bits for receiver 1, and allocate
  \begin{align} \label{eq_DD_corner_smallR0bit}
  k_0 = \frac{R_0}{R^\ast_1}k_1~~ \; \mbox{and} \;  ~~k_2 = \frac{R^\ast_2}{R^\ast_1}k_1,
  \end{align}
  common bits and private bits for receiver 2 respectively. The re-transmission scheme comes as follows.

  \noindent \textbf{Phase 1:} This phase is exactly the same as that of Case I. \\
  \noindent \textbf{Phase 2:} In this phase, we send out private bits for receiver 2 using
  \begin{align} \label{eq_DD_T2}
  n_2 = \frac{k_2}{1-\dl_{12}}
  \end{align}
  time slots, and at the end of this phase, we have
  \begin{align} \label{eq_DDk21}
  k_{2|1} = \frac{(\dl_2-\dl_{12})}{1-\dl_{12}}k_2
  \end{align}
  bits mis-sent to receiver 1.

  \noindent \textbf{Phase 3:}  This phase has three segments.

  \noindent \underline{Phase 3 Segment a}:  Encode the $k_{2|1}$ bits needed for receiver 2 using erasure code at rate $(1-\dl_2)$, and encode
  \begin{align}
  \frac{(1-\dl_1)}{(1-\dl_2)}k_{2|1}
  \end{align}
  bits from $k_{1|2}$ bits needed for receiver 1 at rate $(1-\dl_1)$. Send the XOR of the encoded bits. The total length is
  \begin{align} \label{eq_DD_smallR0T3a}
  n_{3a} = \frac{k_{2|1}}{(1-\dl_2)}
  \end{align}
time slots.

  \noindent \underline{Phase 3 Segment b}: The total length of this segment is
  \begin{align}
  n_{3b} = \frac{1}{(1-\dl_1)}\left( k_{1|2} - \frac{(1-\dl_1)}{(1-\dl_2)}k_{2|1} \right)
  \end{align}
	time slots. We encode the remaining
  \begin{align}
  k_{1|2} - \frac{(1-\dl_1)}{(1-\dl_2)}k_{2|1}
  \end{align}
  bits needed for receiver 1 using an erasure code with  rate $(1-\dl_1)$, and encode
  \begin{align} \label{eq_DD_k03b}
  k_{0,3b} = (1-\dl_2) n_{3b}=\frac{(1-\dl_2)}{(1-\dl_1)} k_{1|2} - k_{2|1}
  \end{align}
  bits from $k_0$ using an erasure code with rate $(1-\delta_2)$. Send the XOR of the encoded bits.

  Note that $n_{3b} \geq 0$ since
  \begin{equation} \label{eq_DD_T3bcheck}
  k_{1|2}\geq \frac{1-\dl_1}{1-\dl_2}k_{2|1}.
  \end{equation}
To see this, note that from \eqref{eq_DD_corner_smallR0} and \eqref{eq_DD_corner_smallR0bit}
  \begin{align}
   \frac{k_1}{k_2}&= \frac{R^*_1}{R^*_2} =  \frac{(1-\dl_1)(\dl_2-\dl_{12})-(\dl_1-\dl_{12})R_0}{(1-\dl_2)(\dl_1-\dl_{12})-(\dl_2-\dl_{12})R_0} \notag \\ & \geq \frac{(1-\dl_1)(\dl_2-\dl_{12})}{(1-\dl_2)(\dl_1-\dl_{12})} \nonumber,
\end{align}
where the inequality holds since $(\dl_1-\dl_{12})R_0 \geq (\dl_2-\dl_{12})R_0$ from $\delta_1<\delta_2$.
Then, we have
  \[ \frac{\dl_1(1-\dl_2)}{(1-\dl_1)(1-\dl_{12})}k_1 \geq \frac{(1-\dl_1)\dl_2}{(1-\dl_2)(1-\dl_{12})}k_2, \]
which turns to \eqref{eq_DD_T3bcheck} by the definitions of $k_{1|2}$ and $k_{2|1}$. We also note that $k_{0,3b} \leq k_0$ as shown in Appendix~\ref{Proof_k03b}.

  \noindent \underline{Phase 3 Segment c}: We encode all $k_0$ common bits using random linear code with length
  \begin{align} \label{eq_DD_T3c}
  n_{3c} = \frac{k_0}{(1-\dl_2)} - n_{3b},
  \end{align}
  and send the encoded bits.

 \begin{remark} \label{Remark:Comparison}
In~\cite{GeorgiadisDetDelayedBC}, authors do not incorporate common bits in Phase~3 Segment~b, and as a result, the corresponding $n_{3c}$ will be longer by $n_{3b}$, resulting in sub-optimal achievable rates when $\delta_1 \neq \delta_2$, see Fig.~\ref{Fig:Region} for an example.   	
\end{remark}	 	

  \noindent {\bf Achievable rate calculation}: At the end of Phase 3, receiver 2 decodes all $k_0$ common bits from $n_{3c}(1-\delta_2)$ bits received in Segment c and
  $
  k_0 - k_{0,3b}
  $
  available side information known from Segment b. The decodability is ensured by  \eqref{eq_DD_k03b} and \eqref{eq_DD_T3c}. Also, receiver 2 has $k_{2|1}$ bits for its private message from Segment a since $k_{1|2}$ mis-sent bits are already known. Together with $(1-\delta_2)n_2$ received bits in Phase 2, all $k_2$ private bits is successfully decoded at receiver 2.

  After Segment c of Phase 3, receiver 1 can first successfully decode $k_{0}$ common bits since
  \begin{align}	\label{eq_DD_T3bcheck1}
	\frac{k_0}{n_{3c}} &= (1-\dl_2) +  \frac{k_{0,3b}}{n_{3c}} \leq 1-\delta_1,
	\end{align}
where the equality comes from \eqref{eq_DD_k03b} and \eqref{eq_DD_T3c} and the proof of the inequality is given in Appendix \ref{Proof_DD_T3bcheck1}. Receiver 1 then removes the interference resulting from the common bits in Segment b of Phase 3 and together with received bits in Segment a and Phase 1, it can decode its intended $k_1$ private bits.

Finally, we calculate the achievable rates. The total communication time is
 \begin{align}
 n &= n_1 + n_2 + n_{3a} + n_{3b} + n_{3c} \nonumber \\
 & \overset{(a)}{=}   n_1 + n_2 + n_{3a} + \frac{k_0}{(1-\dl_2)} \nonumber \\
 & \overset{(b)}{=} \frac{k_1}{1-\dl_{12}} + \frac{k_2}{1-\dl_{12}} + \frac{k_{2|1}}{(1-\dl_2)} + \frac{k_0}{(1-\dl_2)} \nonumber \\
 & \overset{(c)}{=} \frac{k_1}{1-\dl_{12}} + \frac{\frac{R_0}{R^\ast_1}+\frac{R^\ast_2}{R^\ast_1}}{(1-\dl_2)}k_1,
 \end{align}
where (a) is from \eqref{eq_DD_T3c}, (b) is from \eqref{eq_DD_T1}\eqref{eq_DD_T2}\eqref{eq_DD_smallR0T3a}, and (c) is from \eqref{eq_DD_corner_smallR0bit} and \eqref{eq_DDk21}.  Thus, the achievable rate for receiver 1 is
 \begin{align}
 \frac{k_1}{n} = \frac{R^\ast_1}{\frac{R^\ast_1}{1-\dl_{12}} + \frac{R_0+R^\ast_2}{(1-\dl_2)}} = R^\ast_1,
 \end{align}
 where the last equality is valid since the corner point $(R_0, R^\ast_1, R^\ast_2)$ is on the boundary of \eqref{eq_DD_B.1}.  From \eqref{eq_DD_corner_smallR0bit}, we achieve $k_2/n=R^\ast_2$ for private message for receiver 2 and $k_0/n=R_0$ for the common message.

\appendix
\subsection{Impact of increasing the common rate} \label{ImpactofR0}

\noindent {\bf Case I $R_0 > \bar{R}$}: In this case, from (\ref{eq_DD_corner_lareR0}), we have
\begin{align}
R_1^\ast + R_2^\ast + R_0 = (1-\dl_{12}) - \frac{\dl_2-\dl_{12}}{1-\dl_{12}} R_0,
\end{align}
which is a decreasing function in $R_0$ since $\dl_{12} \leq \dl_2$.

\noindent {\bf Case II $R_0 \leq \bar{R}$}: In this case, from (\ref{eq_DD_corner_smallR0}), we have
\begin{align}
& R_1^\ast + R_2^\ast + R_0 \nonumber \\
& = \frac{(1-\dl_1)(\dl_2-\dl_{12}) + (1-\dl_2)(\dl_1-\dl_{12}) - A R_0}{(1-\dl_{12})-\frac{(1-\dl_1)(1-\dl_2)}{(1-\dl_{12})}},
\end{align}
where
\begin{align}
A = \frac{(\dl_1-\dl_{12})(\dl_2-\dl_{12})}{(1-\dl_{12})} \geq 0.
\end{align}
Thus, $\left( R_1^\ast + R_2^\ast + R_0 \right)$ is a decreasing function in $R_0$.

\subsection{Proof of \eqref{eq_hatR1_largeR0}} \label{proof_hatR1_largeR0}

Note that showing \eqref{eq_hatR1_largeR0} is equal to showing
\begin{equation} \label{eq_hatR1_largeR01}
\frac{k_{0,2a}}{n_{2b}} < (\dl_2-\dl_1).
\end{equation}
To do this, from $R_0 \geq \bar{R}$,
we have
\[
  \frac{1}{R_0} - \frac{1}{(1-\dl_2)} < \frac{(\dl_2-\dl_1)}{(1-\dl_2)(\dl_1-\dl_{12})},
\]
where the assumption $\delta_1 \leq \delta_2$ is applied. With $k_0=(R_0/R^*_1)k_1$ and \eqref{eq_DD_corner_lareR0}, the above inequality becomes
\[
	\frac{(\dl_1-\dl_{12})}{(1-\dl_{12})}k_1 < \frac{(\dl_2-\dl_1)}{(1-\dl_2)}k_0,
\]
and together with \eqref{eq_T2a_largeR0}, we have
\[
(1-\dl_1)n_{2a} < \frac{(\dl_2-\dl_1)}{(1-\dl_2)}k_0,
\]
which can be re-organized as
\[
	(1-\dl_2)n_{2a} \leq (\dl_2-\dl_1)\left(\frac{k_0}{(1-\dl_2)}-n_{2a}\right).
\]
Now applying \eqref{eq_T2b_largeR0}, we get $(1-\dl_2)n_{2a} < (\dl_2-\dl_1) n_{2b}$
which validates \eqref{eq_hatR1_largeR01} since $n_{2a}(1-\delta_2)=k_{0,2a}$ by selection.

\subsection{Proof of $k_{0,3b} \leq k_0$} \label{Proof_k03b}

\begin{align}
& ~~~~~\frac{\dl_2-\dl_1}{1-\dl_1} \leq 1 \nonumber \\
& \Rightarrow \frac{(\dl_2-\dl_1)(\dl_1+\dl_2-2\dl_{12}-\dl_1\dl_2+\dl_{12}^2)}{(1-\dl_1)(\dl_1+\dl_2-2\dl_{12}-\dl_1\dl_2+\dl_{12}^2)} \leq 1 \nonumber \\
& \Rightarrow \frac{(1-\dl_1)(\dl_2-\dl_{12})^2-(1-\dl_2)(\dl_1-\dl_{12})^2}{(1-\dl_1)\left( (1-\dl_{12})^2 - (1-\dl_1)(1-\dl_2) \right)} \leq 1 \nonumber \\
& \Rightarrow  \frac{(1-\dl_2)(\dl_1-\dl_{12})R_1^{\ast}}{(1-\dl_1)(1-\dl_{12})R_0}-\frac{(\dl_2-\dl_{12})R_2^{\ast}}{(1-\dl_{12})R_0} \leq 1 \nonumber \\
& \Rightarrow  \frac{1-\dl_2}{1-\dl_1}k_{1|2}-k_{2|1} \leq k_0 \nonumber \\
& \Rightarrow k_{0,3b} \leq k_0.
\end{align}

\subsection{Proof of \eqref{eq_DD_T3bcheck1}} \label{Proof_DD_T3bcheck1}	
 Note that proving \eqref{eq_DD_T3bcheck1} is equal to proving
 \begin{equation} \label{eq_hatR1_smallR01}
 (\dl_2 - \dl_1) \geq \frac{k_{0,3b}}{n_{3c}}.
 \end{equation}
 To do this, we have
 \begin{align}
 (\dl_2-\dl_1)k_0 &=  \frac{(\dl_2-\dl_{12})}{(1-\dl_{12}) - \frac{(1-\dl_1)(1-\dl_2)}{(1-\dl_{12})}} \frac{(1-\dl_1)(\dl_2-\dl_{12})}{(1-\dl_{12})}k_0  \nonumber \\
 &-\frac{(\dl_1-\dl_{12})}{(1-\dl_{12}) - \frac{(1-\dl_1)(1-\dl_2)}{(1-\dl_{12})}} \frac{(1-\dl_2)(\dl_1-\dl_{12})}{(1-\dl_{12})}k_0, \label{eq_DD_T3check2}
 \end{align}
 since
 \begin{align}
 & (\dl_2-\dl_1) \left\{ (1-\dl_{12})^2 - (1-\dl_1)(1-\dl_2) \right\} \notag \\  = & (1-\dl_1)(\dl_2-\dl_{12})^2-(1-\dl_2)(\dl_1-\dl_{12})^2. \notag
 \end{align}
 Together with \eqref{eq_DD_corner_smallR0} and \eqref{eq_DD_corner_smallR0bit}, \eqref{eq_DD_T3check2} comes to
 \[
 (\dl_2-\dl_1)k_0 = \frac{(1-\dl_2)(\dl_1-\dl_{12})}{(1-\dl_{12})}k_1 - \frac{(1-\dl_1)(\dl_2-\dl_{12})}{(1-\dl_{12})}k_2,
 \]
 which implies
 \[
 (\dl_2-\dl_1)k_0 = (1-\dl_2)k_{1|2} - (1-\dl_1)k_{2|1}
 \]
 from \eqref{eq_DDk12} and \eqref{eq_DDk21}. Divide both sides of the above equality with $(1-\dl_2)$, together with the second equality of \eqref{eq_DD_k03b},
 \[
 (\dl_2-\dl_1)\left( \frac{k_0}{(1-\dl_2)}-n_{3b} \right) = (1-\dl_2)n_{3b}.
 \]
Then, from \eqref{eq_DD_T3c} and the first equality of \eqref{eq_DD_k03b}, we have \eqref{eq_hatR1_smallR01}.


\bibliographystyle{ieeetr}
\bibliography{bib_Common}

\end{document}